\documentclass[useAMS,usenatbib]{mn2e}
\usepackage{epsfig}
\usepackage{amsmath}
\usepackage{amssymb}
\newcommand{\abs}[1]{\left|#1\right|}           

\title[Simulation of stellar instabilities 
using domain decomposition]
{Simulation of stellar instabilities with vastly different timescales 
using domain decomposition}
\author[M. Grott et al.]
{M. Grott$^{1}$\thanks{E-mail: mgrott@gwdg.de (MG)},  
S. Chernigovski$^{2}$ and
W.Glatzel$^{1}$\\
$^{1}$Universit\"ats-Sternwarte G\"ottingen, Geismarlandstr .11, 
37073 G\"ottingen, Germany\\
$^{2}$Institut f\"ur Analysis und Numerik, Universit\"at Magdeburg, 
Universit\"atsplatz 2, 39106 Magdeburg, Germany}
\begin{document} 


\pagerange{\pageref{firstpage}--\pageref{lastpage}} \pubyear{2002}

\maketitle

\label{firstpage}

\begin{abstract}
Strange mode instabilities in the envelopes of massive stars lead to
shock waves, which can oscillate on a much shorter timescale than that 
associated with the primary instability. The phenomenon is studied by
direct numerical simulation using a, with respect to time, 
implicit Lagrangian scheme, which allows for the variation by several 
orders of magnitude of the dependent variables. 
The timestep for the simulation
of the system is reduced appreciably by the shock oscillations 
and prevents its long term study.
A procedure based on domain decomposition is proposed to surmount the 
difficulty of vastly different timescales in various regions of the stellar
envelope and thus to enable the desired long term simulations. Criteria
for domain decomposition are derived and the proper treatment of the resulting
inner boundaries is discussed. Tests of the approach are presented and 
its viability is demonstrated by application to a model for the star
P Cygni. In this investigation primarily the feasibility of
domain decomposition for the problem considered is studied. We
intend to use the results as the basis of an extension to two 
dimensional simulations. 
\end{abstract}


\section{INTRODUCTION}

Sufficiently luminous objects, such as massive stars,
are known to suffer from strange mode instabilities with growth 
rates in the dynamical range \citep{KFG93,GK93}. 
The boundary of the domain in the Hertzsprung-Russel diagram (HRD) 
above which all stellar models are unstable - irrespective of their 
metallicity -, coincides with the observed Humphreys-Davidson (HD) 
limit \citep{HD79}. Moreover, the range of unstable models covers the 
stellar parameters for which the LBV (luminous blue variable) 
phenomenon is observed (for a review see \citet{HD94}).

The high growth rates of the instabilities indicate a connection to 
the observed mass loss of the corresponding objects. To verify this suggestion,
simulations of their evolution into the non linear regime have been performed.
In fact, for selected models \citet{GKCF99} found 
the velocity amplitude to exceed
the escape velocity (see, however, \citet{DG00}).

To identify a possible connection between non linear pulsations and outbursts in 
luminous blue variables \citet{GGC02} have studied the evolution of 
an initial model located in the HRD well above the HD limit.
In this study, the shocks formed in the non linear regime are captured 
by the H-ionization zone after a few pulsation periods. 
These captured shocks start to oscillate rapidly with
periods of the order of the sound travel time across the H-ionisation zone, 
while its mean position changes on the dynamical timescale of the primary,
strange mode instability. \citet{GGC02} have 
shown, that this shock front 
oscillation is of physical origin and therefore must not be disregarded.
In particular, the phenomenon should not be eliminated by 
increasing the artificial viscosity. We note that the representation of the
phenomenon requires the correct treatment of extreme gradients of the dependent
variables, implying their variation by several orders of magnitude. It
is achieved by use of a, with respect to time implicit Lagrangian scheme.

The rapid shock oscillations, which are
confined to a narrow region in the vicinity of the shock front,
require an inhibitively small
timestep and thus prevent long term simulations. In the present paper we
propose an approach based on domain decomposition to surmount the 
difficulty of vastly different timescales in various regions of the stellar
envelope and thus to enable the desired long term simulations. 
In this procedure the various domains within the envelope are to be
treated separately and according to their intrinsic timescales.
We expect this decomposition to speed up the calculations considerably.
An even higher speedup will be achieved when applying domain decomposition 
to two dimensional simulations. In this sense, the present investigation may be 
regarded as a preliminary study for decomposition in two dimensions.

The basic equations and assumptions are introduced in Section \ref{basic}.
The domain decomposition approach is discussed in detail in Section \ref{domain},
including a derivation of criteria for domain decomposition and an investigation
of the proper treatment of the resulting
inner boundaries. Moreover, tests of the approach are presented there. 
Its viability is demonstrated by application to a model of the star 
P Cygni in section \ref{PCygni}.  
Our conclusions follow.

\section{BASIC EQUATIONS AND ASSUMPTIONS}
\label{basic}

The evolution of instabilities of a stellar envelope is followed into the non-linear
regime assuming spherical symmetry and adopting a Lagrangian description, 
i.e. the independent
variables are the time t and the mass $M_r$ inside a sphere of radius $r$. 
The 
equations to be solved (see, e.g., \citet{C80}) are given by
mass conservation,
\begin{eqnarray}
  \frac{\partial r^3}{\partial M_r} -\frac{3}{4\pi\rho}&=&0
  \label{nl1}
\end{eqnarray}
momentum conservation,
\begin{eqnarray}
  \frac{\partial^2 r}{\partial t^2} +4\pi r^2 
  \frac{\partial p}{\partial M_r} +\frac{GM_r}{r^2}&=&0
  \label{nl2}
\end{eqnarray}
energy conservation,
\begin{eqnarray}
  \frac{\partial L}{\partial M_r}- \epsilon - \frac{p}{\rho^2}
  \frac{\partial \rho}{\partial t}+\frac{\partial E}{\partial t} &=&0
  \label{nl3}
\end{eqnarray}
and the diffusion equation for energy transport,
\begin{eqnarray}
  \frac{\partial T}{\partial M_r}-\frac{3\kappa(L-L_{conv})}{64\pi^2acr^4T^3} &=&0 
  \label{nl4}
\end{eqnarray}
where $\rho$, $p$, $T$, $L$, and $E$ denote density, pressure, temperature, 
luminosity and specific internal energy, respectively. $a$ is the Stefan-Boltzmann
constant, $G$ the gravitational constant and $c$ the speed of light. 
We emphasise that $\partial\over \partial t$ is the substantial time 
derivative. 
For the opacities $\kappa$, the latest versions of the 
OPAL tables \citep{IRW92,RI92} have been used.  
Convection is treated in the standard
frozen in approximation (see \citet{BK65}), 
i.e. the convective luminostity $L_{conv}$ is kept constant
and equal to the luminosity of the hydrostatic initial model. 
Since the instabilities are localized in the outer envelope, the evolution of 
the core can be neglected. Its properties are taken into account by imposing 
time independent boundary conditions 
(e.g. by prescribing luminosity, [vanishing] velocity $v$ and [constant] radius) 
at the bottom of the envelope.
For the envelope model, the nuclear energy generation rate $\epsilon$ 
vanishes.
At the outer boundary, the gradient of heat sources is required to vanish ($F$ is 
the heat flux):
\begin{eqnarray}
  {\rm{grad}}({\rm{div}} F)=0 
\end{eqnarray}
This boundary condition implies (by using equations \ref{nl1},\ref{nl3} 
and \ref{nl4}) boundary values for the 
temperature $T$ and pressure $p$.
It is chosen to ensure that outgoing shocks pass through the boundary without
reflection.
The set of boundary conditions prescribing values for the velocity and luminosity
at the bottom and values for the pressure and temperature at the top of the
envelope will be denoted by $(v,L)(p,T)$ in the following. 

The numerical code relies on 
a Lagrangian, with respect to time implicit, fully conservative 
difference scheme  proposed by 
\cite{FR68} and \cite{SP69}. As a consistent extension the two dimensional
version of the code currently under development (also implicit and 
Lagrangian) is based on the method of support operators originally
suggested by \cite{ST81} and \cite{AG82}.
For a thorough description of the method of support operators
we refer to \citet{Sh96}.
To handle shock waves, artificial viscosity 
is used. Concerning tests of the code, we adopted the same criteria as \citet{GKCF99}.
We emphasise, that conservativity of the numerical scheme is of fundamental importance
when simulating instabilities in a stellar envelope. Considering the distribution 
of internal, gravitational and kinetic energy we find, that
the kinetic energy can be smaller than, e.g., the gravitational energy
by several orders of magnitude.
Appropriate simulations of stellar instabilities  require the correct
representation of the kinetic energy, and therefore energy conservation 
with high accuracy is indispensable. The difference scheme adopted here
guarantees energy conservation.

\section{DOMAIN DECOMPOSITION}
\label{domain}

\subsection{Motivation for domain decomposition}

\begin{figure*}
  \epsfig{file=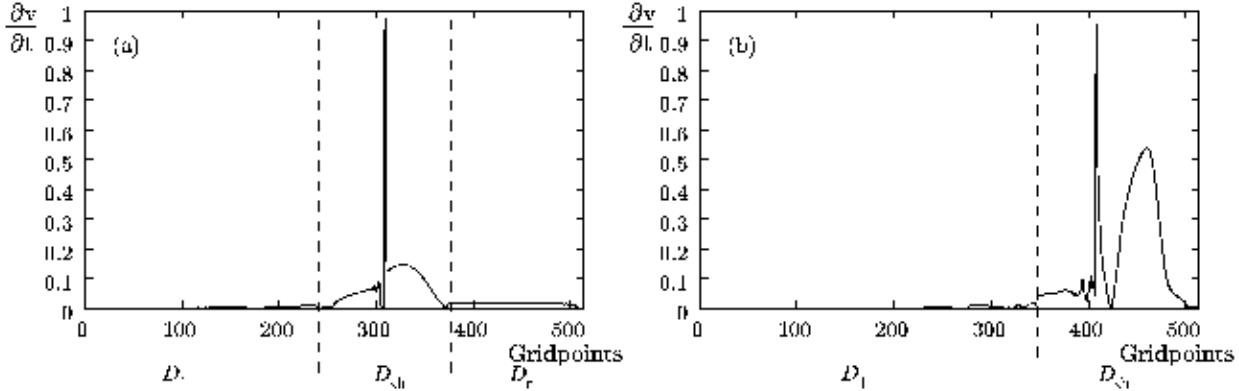,width=16.8cm}
  \caption{Normalized time derivatives of the velocity $v$ as a function of 
    gridpoint, for two typical states of the system. 
    Decomposition into three (a) and two (b)
    subdomains ($\mathcal{D}_{l}$, $\mathcal{D}_{sh}$, $\mathcal{D}_{r}$)
    as indicated by the dashed lines is suggested.}
  \label{dt} 
\end{figure*}

The stellar envelope model representing a massive star with mass $M=50M_\odot$, 
effective temperature $T_{\rm{eff}}=10^4K$ and luminosity 
$L=1.17\cdot 10^6 L_\odot$, which was considered by \citet{GGC02}, suffers from 
strange-mode-instabilities. These cause pulsations with
velocity amplitudes of 0.5 $v_{\rm{esc}}$ and inflate the envelope to $2.5$ initial
radii. After several pulsation periods a shock front 
is captured in the H-ionization zone. The latter is prone to secondary instabilities
and oscillates on very short timescales connected to the sound travel time 
across the front. These instabilities are caused by the stratification, but not
driven by buoyancy \citep{GGC02}.
Resolving the rapid oscillations of the shock front reduces the timestep to very
small values. This is computationally 
extremely expensive and effectively inhibits the desired long term study of 
the system.

In order to enable the treatment of the problem the integration interval is decomposed
such that the small and quickly varying shock region is integrated with small time
steps and the remaining major part of the envelope is calculated using large 
time steps. This strategy of domain decomposition  is common in computational 
fluid dynamics (see, e.g., \cite{W99} and \cite{WZ00}). For the treatment of
interfaces between the various domains and the associated inner boundary 
conditions for purely hyperbolic systems, we refer the reader to these publications. 
We stress, however, some fundamental differences to the
previous studies: One of them concerns the different character of the 
system of equations.
So far, only purely hyperbolic systems have been considered, whereas we 
apply domain decomposition to a composite
system of hyperbolic and parabolic equations. 
Moreover, we adopt a numerical
scheme which is implicit in time. Again, this is in contrast to previous studies
which use explicit schemes.

\subsection{Criterion for domain decomposition}
\label{criterion}
In this section a criterion for the proper choice of the boundaries of the various
domains evolving on different timescales and therefore treated with different
time steps will be presented.  
Considering the time derivatives of various physical quantities the velocity 
$v$ is found to vary most rapidly and therefore determines the time step.

Figure \ref{dt} shows the time derivatives of the velocity for two typical
states of the system, where Figure \ref{dt}.a corresponds to rapid shock oscillations
around gridpoint 310. In this case, domain decomposition as indicated by the 
dashed lines is suggested. 
Figure \ref{dt}.b represents a situation,
in which the outer envelope is collapsing onto the shock (at
gridpoint 410). Rapid variations are now also found above the shock.
Accordingly, decomposition into two domains (as indicated by the dashed line)
seems appropriate. Depending on the state of the system, we therefore need 
to split the domain of integration into two or three subdomains. 

The size of the various domains is determined by comparing 
the time derivatives of velocity
and temperature with the corresponding derivatives on the shock front. Therefore,
as a first step, the position of the shock front has to be determined. For the 
models studied, the latter is defined by the maximum temperature gradient.
The boundaries of the shock zone  are then defined by
the requirement, that the time derivative at their position corresponds to 
a given fraction $1/k$ of the time derivative at the shock.
In other words, the set of gridpoints $\mathcal{D}_{sh}$ belonging to the shock zone 
may be characterized by 
\begin{eqnarray}
  \mathcal{D}_{sh}=\left\{  
    n: \abs{\frac{\partial v}{\partial t}}_{n}
    > \frac{1}{k}\abs{\frac{\partial v}{\partial t}}_{sh} \wedge 
\abs{\frac{\partial T}{\partial t}}_{n}
    > \frac{1}{k}\abs{\frac{\partial T}{\partial t}}_{sh} \right\}   
\label{k}
\end{eqnarray}
where  $n$ denotes the number of the gridpoint and derivatives with index $sh$
correspond to their maximum values found in the shock region.
Accordingly, the zones below and above the shock are determined by
\begin{eqnarray}
  \mathcal{D}_l &=& \left\{n:n\le \mbox{min}(\mathcal{D}_{sh})\right\} \\ 
  \mathcal{D}_r &=& \left\{n:n\ge \mbox{max}(\mathcal{D}_{sh})\right\}. 
\end{eqnarray} 
The parameter $k$ corresponds to the ratio of timesteps in $\mathcal{D}_{l,r}$
and $\mathcal{D}_{sh}$, which favours large values of $k$. On the other hand,
the size of $\mathcal{D}_{sh}$ increases with $k$, suggesting smaller values.
Therefore optimum results will be obtained for a mean choice.
For the considered model a value of $k=15$ turned out to be satisfactory.
If the size of the domains  $\mathcal{D}_l$ or $\mathcal{D}_r$ drops below a
given value, they are considered to be part of the shock zone, and we arrive at
a decomposition into two subdomains. For the model considered   
the whole grid consists of 512 gridpoints and the zones have a minimum size
of 64 gridpoints. 

After decomposition, the quickly varying region $\mathcal{D}_{sh}$
is integrated first with timesteps $\tau_1,\tau_2,\ldots,\tau_n$. Then
the domains $\mathcal{D}_l$ and/or $\mathcal{D}_{r}$ are integrated with the timestep
$\tau=\tau_1+\tau_2+\cdots+\tau_n$. The decomposition implies artificial boundaries
and boundary conditions between the domains $\mathcal{D}_l$ and $\mathcal{D}_{sh}$ and 
$\mathcal{D}_r$ and $\mathcal{D}_{sh}$, respectively. 
An inconsistency is introduced, if in a first approach the explicit inner 
boundary conditions are either kept fixed, or linearly extrapolated in time.
Both cases are contained in the following extrapolation presciption:
\begin{eqnarray}
  Y_{bound}&=&\alpha_Y\left(\frac{(Y-Y_{old})}
    {\tau_{old}}\cdot{\tau}'+Y
  \right)
  +(1-\alpha_Y)\cdot Y
  \label{extrap}
\end{eqnarray}
$Y$ stands for $p$, $T$, $v$, $L$, and $\alpha_Y\in [0,1]$
is a free extrapolation parameter, which may be chosen independently for each variable.
$Y$ refers to the current, $Y_{old}$ to the previous value of the variable. 
$\tau_{old}$ is the last, $\tau'$ the current timestep.
How the inconsistency may be treated will be 
discussed in the following sections.

Once the integration of all domains has been performed, one subsequent timestep is
done without decomposition. We shall refer to it as the relaxation timestep.
Relaxation prevents accumulation of residual errors.

\subsection{The iteration procedure}
\label{iteration}

A mathematically consistent way of integrating the different domains, which solves
the problem of artificial fixed boundary conditions is the
following iterative procedure:
\begin{enumerate}
\item \label{i} The computation is started at time $t$ and 
  $D_l$, $D_{sh}$ and $D_{r}$ are integrated
  with fixed boundary values $(v,L)_t$ and $(p,T)_t$ for each domain.
  The subscript $t$ denotes values at time $t$.
\item New boundary values  $(v,L)_{t+\tau}$ and $(p,T)_{t+\tau}$ are obtained
  by integration of the adjacent domain.
  With these boundary values the integration \ref{i} is repeated.
\end{enumerate}
This iterative procedure implies implicit boundary conditions and
sucessively eliminates the errors introduced by the artificial 
boundaries. 4-5 iteration cycles produce satisfactory results.
However, this approach is computationally even more 
expensive than integrating the entire domain with small timesteps 
$\tau_1,\ldots,\tau_n$. Concerning the present problem, it is  therefore not relevant 
and has been applied only for comparison with other methods discussed below. 
However, for the corresponding two dimensional problem the computational 
effort might be significantly reduced by the procedure as the inversion of 
the most ill-conditioned matrix occurring there is an $N^2$-process.

\subsection{Overlapping domains}

\begin{figure*}
  \epsfig{file=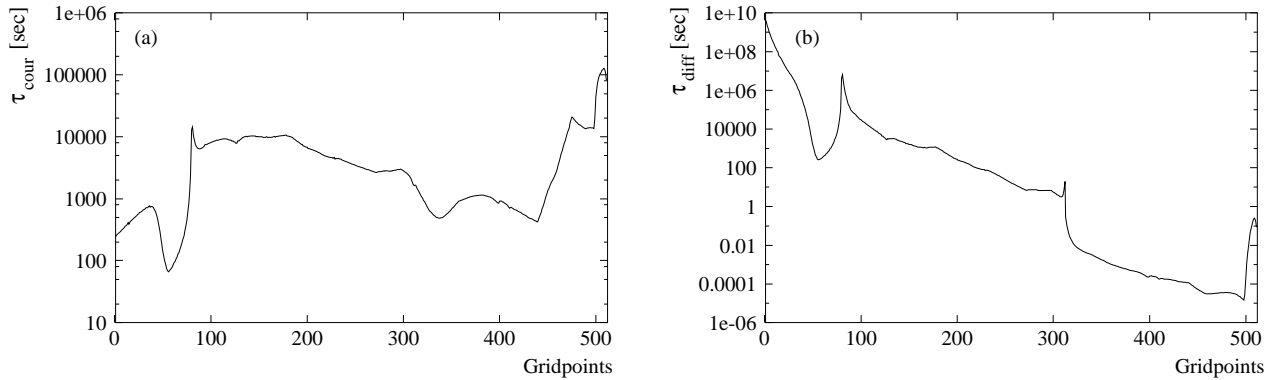,width=16.8cm}
  \caption{Snapshot of the sound travel time $\tau_{\rm{cour}}$ (a) and 
    the diffusion time 
    $\tau_{\rm{diff}}$ (b) across a cell as a function of gridpoint.}
  \label{tau}
\end{figure*}
One method to reduce the error caused by the fixed inner boundary conditions consists
of using overlapping domains of computation. We emphasise that the system considered
here is composed of  hyperbolic and parabolic differential 
equations. 
We first consider the hyperbolic part, i.e., the mechanical equations. 
Any errors or perturbations in this part
produced at the boundaries of the domains propagate with
finite speed into the domain of computation
along the characteristics of the equation. 
For our set of equations, perturbations propagate
with velocities $v\pm c_s$, where $c_s$ denotes the speed of sound.
If the domains overlap such that 
the time for error propagation across the overlap is larger than the integration 
timestep $\tau$, we may discard the flawed values close to the boundary and 
keep only the correct values for the subsequent timestep. 
This condition implies 
\begin{eqnarray}
  \tau < \sum_{i} \frac{l_i}{v_i+c_{s_i}}
\label{hyperbolic}
\end{eqnarray}
where the sum of individual sound travel times is to be taken over 
all overlapping cells. $l_i$ denotes the thickness of cell $i$.

Concerning the parabolic part of equations \ref{nl1}-\ref{nl4}, i.e. the
diffusion equation for energy transport, perturbations travel 
across the grid with infinite speed. Therefore, the procedure suggested 
for the hyperbolic
part of the equations can in principle not be carried over to the parabolic part. 
Rather physical quantities change on the diffusion timescale, which
 is given by $\tau_{\rm{diff}}=
  l^2\rho c_p\frac{3\kappa\rho}{4 a c T^3}$ ($c_p$ is the specific heat at constant
pressure). Consequently, we expect the effect of the
perturbations to be small far from the
boundary if the timestep $\tau$ is sufficiently small, i.e. :
\begin{eqnarray}
  \tau <\sum_{i}
  l_i^2\rho_i c_{p_i}\frac{3\kappa_i\rho_i}{4 a c T_i^3}
\label{parabolic}
\end{eqnarray}
where the sum of individual diffusion times again extends over all overlapping
cells.

In Figure \ref{tau} a snapshot of the sound travel time  
(Figure \ref{tau}.a) and diffusion time (Figure \ref{tau}.b) across a cell
is given as a function
of gridpoint. The sound travel time across the overlapping region is larger than 
a typical timestep (approximately 1 per cent of the global free fall time) 
for an overlap of $\sim 8$ cells (condition \ref{hyperbolic}).
However, while the diffusion time across the overlapping region is bigger than
the timestep even for a small overlap in the bottom part of the envelope,   
condition \ref{parabolic} cannot be satisfied in the outer envelope for 
a reasonable number of overlapping cells.  
The latter is due to the small heat capacity there (implying the ratio of 
thermal and dynamical timescales to be small)  and in accordance with the
validity of the non-adiabatic-reversible approximation (NAR-approximation),
which has been shown for this particular stellar model by 
\citet{GGC02}. How satisfactory results may be obtained even if condition
\ref{parabolic} is not satisfied will be discussed in the following section.

\subsection{Inner boundary conditions}
\label{boundaries}
\begin{figure*}
  \epsfig{file=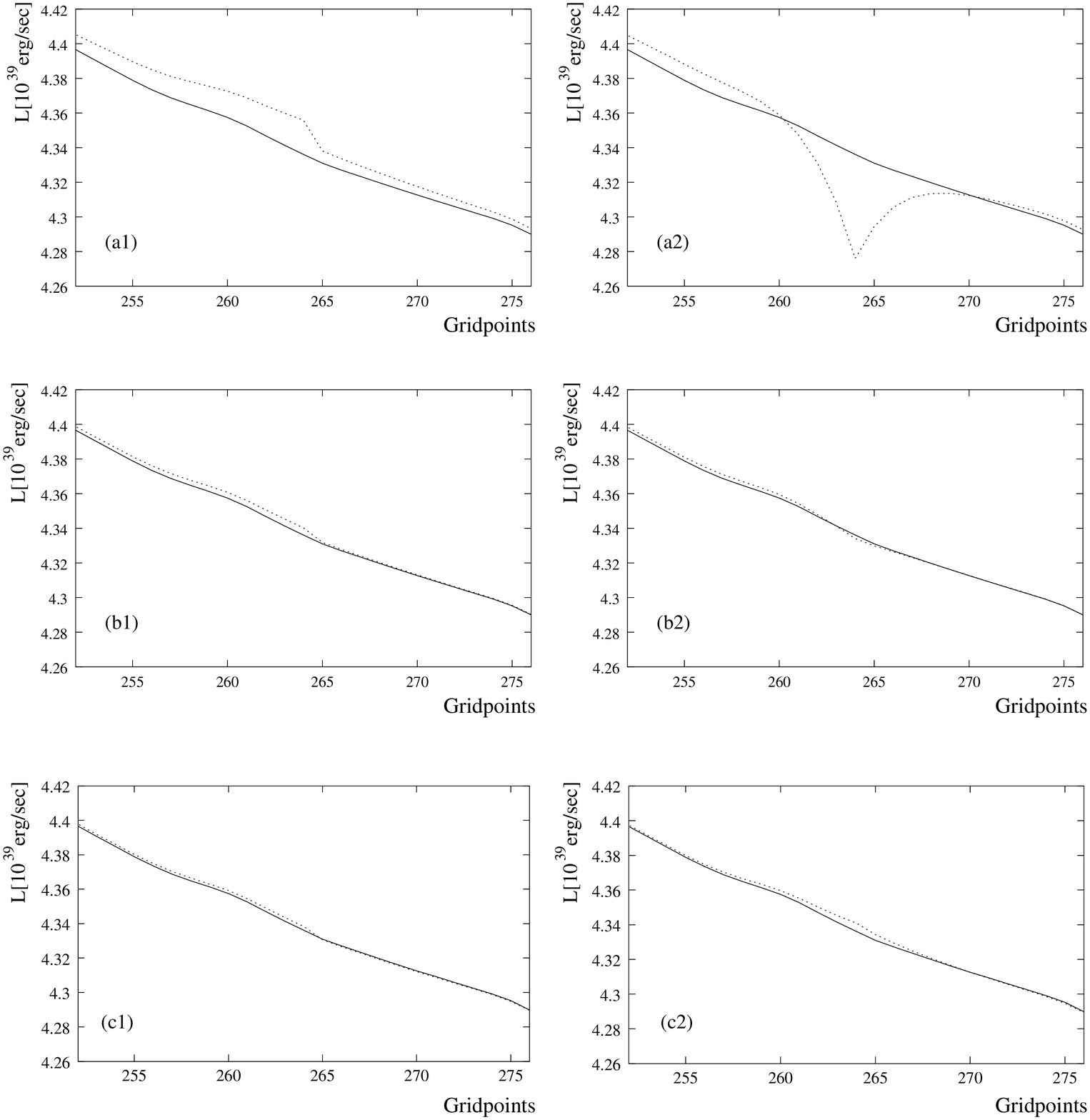,width=16.8cm}
  \caption{The luminosity as a function of gridpoint around
    the boundary between $\mathcal{D}_l$ and $\mathcal{D}_{sh}$ at gridpoint 264.
    Results obtained without decomposition are shown as solid lines. Dotted lines
    in the left and right columns correspond to decomposition without (1) and with (2)
    relaxation, respectively. 
    $(v,L)(p,T)$ and  $(v,L)(v,L)$  boundary conditions have been used in (a) and (b),
    respectively, the iteration procedure (4 cycles) has been applied in (c) with
    the $(v,L)(p,T)$ boundary condition.}
  \label{boundary}
\end{figure*}

On the basis of the domain decomposition procedure described above, 
various inner boundary conditions and their consequences will be investigated
in this section. 
In any case, overlapping domains have been used. The tests presented here 
have been performed at various times with similar results. 
Therefore the results may be regarded to be independent of the particular
state of the system. 
In the tests, the luminosity $L$ turned out to be the most 
sensitive quantity concerning the errors introduced by the inner boundaries.
This is due to the fact, that in our difference scheme it is only accurate to 
first order.  Therefore $L$ is required to be reproduced satisfactorily
compared to the results of the approach without decomposition.

In Figure \ref{boundary} the luminosity is given as a function of gridpoint around
the boundary between $\mathcal{D}_l$ and $\mathcal{D}_{sh}$, which is more sensitive
than the boundary between $\mathcal{D}_{sh}$ and $\mathcal{D}_{r}$
with respect to the decomposition procedure. For comparison, results obtained without
decomposition are shown as solid lines. Dotted lines correspond to results with
decomposition, where the left and right columns illustrate those before and 
after relaxation, respectively.

\begin{enumerate}
\item Figure \ref{boundary}.a shows the result obtained using the 
  $(v,L)(p,T)$ boundary condition, i.e., velocity and luminosity were prescribed
  at the left, and temperature and pressure at the right boundary, respectively.
  This condition implies a discontinuity of the luminosity (a1) and leads
  to an unacceptable relative error ($\sim$ 15 per cent) after relaxation (a2).
\item Figure \ref{boundary}.b shows the result obtained using the 
  $(v,L)(v,L)$ boundary condition. The latter is motivated by the discontinuity of
  the luminosity for the previously discussed $(v,L)(p,T)$ boundary condition.
  Rather than the considerable discontinuity of $L$ we expect a 
  more tolerable discontinuity of its derivative for the $(v,L)(v,L)$ boundary 
  condition. In fact, the results given in 
  figure \ref{boundary}.b1 are satisfactory. After relaxation 
  the relative error in the luminosity is of the order of $10^{-3}$
  (figure \ref{boundary}.b2). However, in this case the error of $L$ directly
  enters the boundary conditions for the subsequent timestep 
  and leads to an accumulation of the error at the interface.
  This problem can be removed partially by switching between different interfaces
  for subsequent timesteps. The remaining error can then spread sufficently
  and does not influence the further integration significantly.
\item Figure \ref{boundary}.c shows the result obtained using the iteration
  procedure described in section \ref{iteration}. After four iteration cycles 
  it is comparable to that obtained with the 
  $(v,L)(v,L)$ boundary condition. 
  By performing more iteration cycles the accuracy could be improved even more.
  However, the iteration procedure is
  computationally much more expensive than the alternatives discussed
  and therefore of no practical use.
\end{enumerate}

We thus conclude, that using the domain decomposition procedure together with
overlapping domains and $(v,L)(v,L)$ boundary conditions, and switching
between different interfaces, yields satisfactory results
at low computational cost.

\subsection{Validation of the domain decomposition procedure}

\begin{figure*}
  \epsfig{file=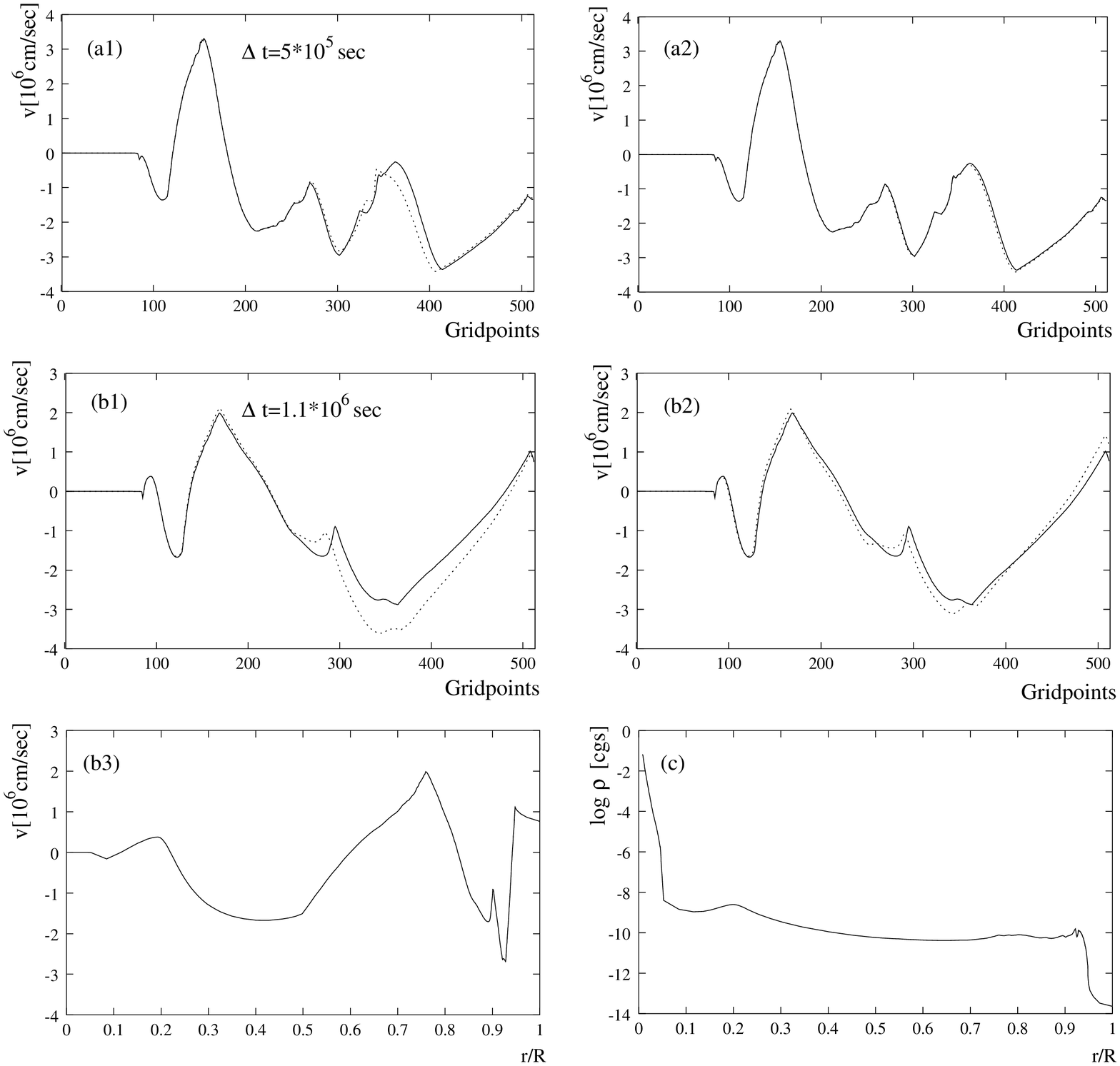,width=16.8cm}
  \caption{The velocity as a function of gridpoint with (dashed line) and without
    (solid line) domain decomposition
    after $5\cdot 10^5$ sec (a1) and $1.1\cdot 10^6$ sec 
    (b1) of simulated time. All simulations start
    at $t=6.47\cdot 10^7$ sec (after the formation of the shock front) with the
    same initial model. In Figures (a2) and (b2) velocities obtained with 
    decomposition are taken $10^4$ sec (a2) and $5\cdot 10^4$ sec (b2) later than their 
    counterparts without decomposition, respectively. This physically 
    irrelevant phaseshift reduces
    the differences between the two approaches significantly. For comparison,
    the velocity and the density after $1.1\cdot 10^6$ sec 
    of simulated time are presented as a function of relative radius  
    in figures (b3) and (c), respectively (without decomposition).}
  \label{time}
\end{figure*} 
\begin{figure*}
  \epsfig{file=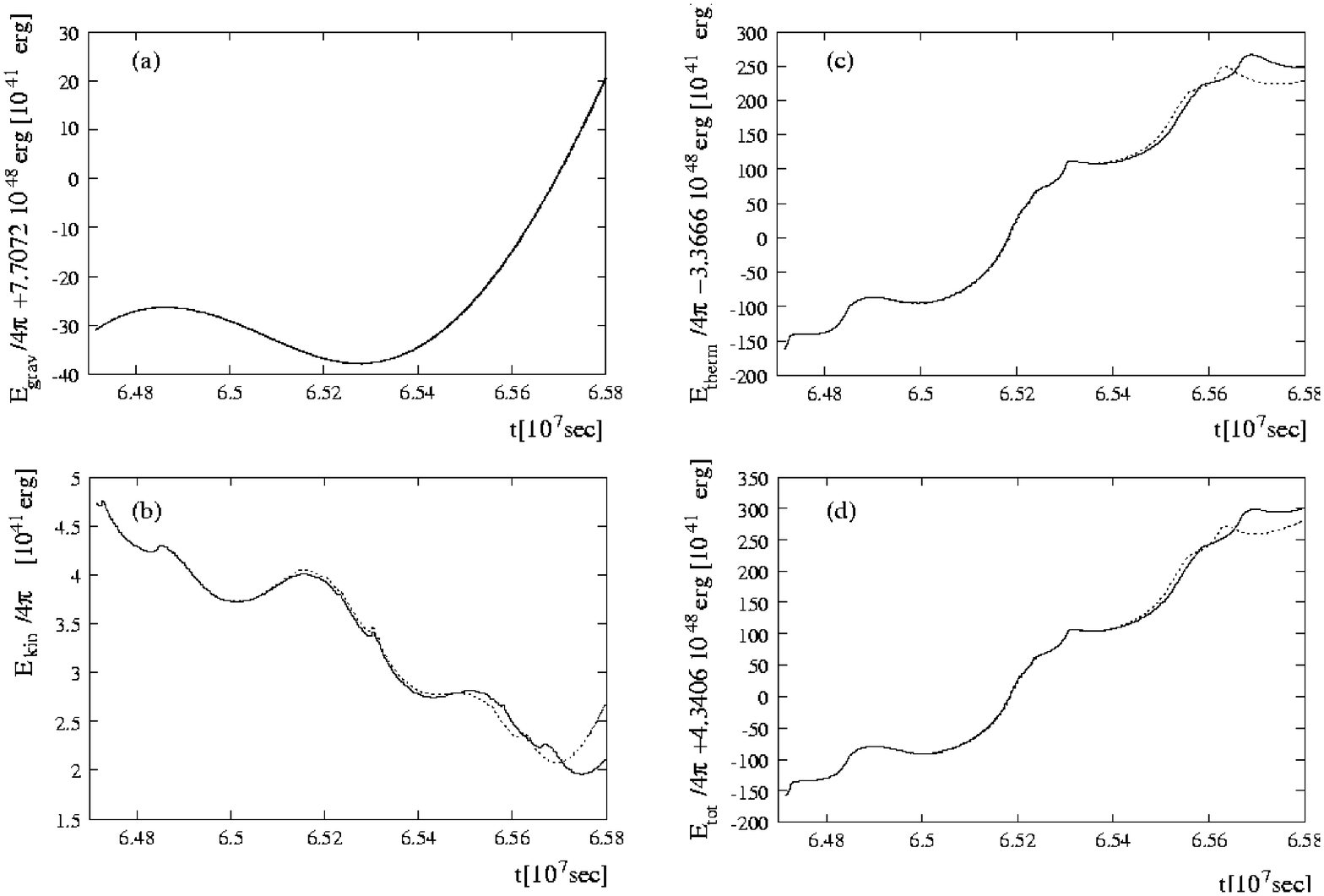,width=16.8cm}
  \caption{Gravitational (a), kinetic (b), thermal (c) and total (d)
    energy as a function of time. Solid and dashed lines 
    correspond to simulations without and with domain decomposition, respectively.}
  \label{energy}  
\end{figure*}
The domain decomposition procedure with overlapping domains and  
$(v,L)(v,L)$ boundary conditions has been compared for validation
with the original approach using no decomposition. 
The comparison starts at $6.47\cdot 10^7$ sec, i.e., well after
the formation of the shock and the associated instability, and extends to
$6.58\cdot 10^7$ sec. The 
typical periods of unstable strange modes driving the pulsations of the
star are of the order of $5\cdot 10^6$ sec, whereas the modes carrying
the stratification instabilities
of the shock front have periods of $\sim 10^5$ sec. Therefore, the test covers
approximately 0.2 periods of the overall envelope pulsations and 10 periods of shock
oscillations. 

Convergence and error control is done using the following 
criterion based on a $l^2$-norm:
\begin{eqnarray}
\frac{1}{N}\left( \sum_{i} f_i^2 \right)^{\frac{1}{2}} < E
\label{l2}
\end{eqnarray}
where $f_i$ denotes the relative error of a physical
quantity in gridpoint or cell $i$ and $N$ the total number of 
gridpoints. $E$ is the prescribed error bound and the sum extends over all gridpoints.
$f_i^2$ contains the weight-function of the $l^2$-norm which is chosen to be 
proportional to the mass of the corresponding cell, $f_i^2\propto m_i$.
Thus the various regions of the star contribute to the error in a different way
and domain decomposition using the same error bound in all domains will result in 
different accuracies for the various domains and compared with the approach 
without domain decomposition. Accordingly the error bound has to be adapted to
the domains contribution to the numerical error. On the other hand, as an identical
error control cannot be guaranteed, results with and without decomposition are
expected to differ slightly.  
 
In the following figures \ref{time} and \ref{energy}, solid lines correspond to  the
approach without decompostion, dotted lines refer to decomposition.
Figure \ref{time} shows the velocity $v$  as a function of gridpoint
after $5\cdot 10^5$ sec 
(figure \ref{time}.a1) and $1.1\cdot 10^6$ sec
(figure \ref{time}.b1) of simulated time, respectively. 
For comparison,
the velocity and the density after $1.1\cdot 10^6$ sec 
of simulated time are also presented as a function of relative radius  
in figures (b3) and (c), respectively (without decomposition).
The shock is located at $r/R\approx 0.95$ and resolved by approximately 150
gridpoints. The high resolution of the shock zone is necessary to 
represent its oscillations (\cite{GGC02}).

Considering $5\cdot 10^5$ sec of simulated time, domain 
decomposition yields  excellent agreement with the original 
approach up to the position of the shock front (at grid point 320). 
Around the shock front
the results differ slightly, whereas above it the 
agreement between the two approaches is again satisfying. 
The agreement may be found to be even better, if a phaseshift in time 
of about $10^4$ sec between the results
discussed is taken into account, i.e., if results of the original approach 
are compared with those obtained 
$10^4$ sec later with decompostion (figure \ref{time}.a2). 
We emphasise, that the time interval corresponds to five oscillation cycles
of the shock front instability, and implies a phaseshift of only $\frac{\pi}{5}$.
With respect to the fact that the phaseshift is physically irrelevant, we thus
regard the agreement as fully satisfying.

After $1.1\cdot 10^6$ sec of simulated time the differences become more pronounced,
in particular in the vicinity of the shock (now at gridpoint 300). Similar to 
the previous discussion, however, including a 
suitable phaseshift of $5\cdot 10^4$ sec a reasonable agreement may be achieved
(Figure \ref{time}.b2).

Figure \ref{energy} shows the gravitational (Figure \ref{energy}.a), 
kinetic (Figure \ref{energy}.b),
 thermal (Figure \ref{energy}.c) and total energy (Figure \ref{energy}.d) 
of the envelope as a function of time. 
We note that the kinetic energy given in figure \ref{energy}.b is
seven orders of magnitude smaller than either the gravitational or 
thermal energy. It may be referred to as the energy of the pulsation and is
therefore of central interest in the present context. Its variation 
with time is one order of magnitude smaller than that of the gravitational 
energy, which itself is one order of magnitude smaller than that of
the thermal energy. Therefore the variation of the total energy is 
dominated by the thermal energy. 

Domain decomposition reproduces the gravitational energy perfectly.
This can be attributed to the use of overlapping domains, since without them
considerable disagreement is found. The latter consists of oscillations of the 
solution around the curve given in Figure \ref{energy}.a. 
With respect to kinetic (Figure \ref{energy}.b), thermal (\ref{energy}.c)
and total energy (\ref{energy}.d), the agreement is satisfactory 
up to the time $6.55\cdot 10^7$ sec. The deviations at later times can 
be partially attributed to the phase shift discussed above. 

To summarise, our comparisons prove the domain 
decomposition procedure to provide reliable and satisfactory results.
We emphasise that domain decomposition violates the conservativity otherwise
inherent in the numerical scheme. How this violation of conservativity contributes
to the discrepancies discussed is an open question.

\subsection{Speed-Up of the calculations}

In order to estimate the speed-up achieved by the domain decomposition 
procedure, we 
assume that the number of iterations $I$ needed to solve the implicit 
equations with prescribed accuracy
is not changed by decomposition, i.e., for convergence $I$ iterations are
required in the shock region $\mathcal{D}_{sh}$ integrated with timestep
$\tau$ and the same number of iterations $I$ are needed to integrate the 
domains $\mathcal{D}_{l}$ and
$\mathcal{D}_{r}$ with time step $k\cdot\tau$. Moreover, we assume
that $I$ iterations are required to integrate with timestep $\tau$
and without decomposition.

We denote the number of gridpoints by $N$ and the size 
of the domains $\mathcal{D}_{l}$, $\mathcal{D}_{sh}$
and $\mathcal{D}_{r}$ by $N_l$, $N_{sh}$ and $N_r$, respectively. 
($N_l$, $N_{sh}$ and $N_r$ include the overlap.)
Integrating $k$ timesteps $\tau$ without decomposition then requires
\begin{eqnarray}
O_{1}= k\cdot N \cdot I
\end{eqnarray}
operations. Decomposing the grid into two or three
domains, we need for the integration of the same time interval
\begin{eqnarray}
O_{2}&=& N_l \cdot I +
k\cdot N_{sh}\cdot I \quad \mbox{and}\\
O_{3}&=& (N_l+N_r) \cdot I +
k\cdot N_{sh}\cdot I
\end{eqnarray}
operations, respectively. Thus we expect a speed-up of the calculations by
decomposition by a factor of
\begin{eqnarray}
s_{2}&=&\frac{O_{1}}
{O_{2}}= 
\frac{k N}{N_l+kN_{sh}}
\quad \mbox{and}\\
s_{3}&=&\frac{O_{1}}
{O_{3}}
=\frac{k N}{(N_l+N_r)+kN_{sh}},
\end{eqnarray}
respectively.
It essentially depends on $N_{sh}$
and $k$. For large $k$ it is approximately given by $N/N_{sh}$. 
However, for large $k$, $N_{sh}$ increases too (see section \ref{criterion}).
 A comparison between 
the estimated and measured speed-up is presented in Table \ref{speed}.
For the hypothetical case of large $k$, $N_{sh}=16$ and an overlap of $8$ gridpoints
we obtain a speed-up by a factor of $16$. This situation could in principle be 
realised for shock oscillations with smooth spatial structure which can be 
represented by a small number of gridpoints. This example demonstrates the
power of the method when applied to a suitable situation.

\begin{table}              
\begin{tabular}{ccccc}   
        &          &     &    Estimated & Measured\\
Domains & Size of $\mathcal{D}_{sh}$& 
Overlap& Speed-Up& Speed-Up\\
\hline
 3  & 160 & 16 & 2.75 & 2.42 \\
 3  & 160 & 32 & 2.72 & 2.36 \\
 3  & 192 & 16 & 2.34 & 2.15 \\
 3  & 192 & 32 & 2.32 & 2.14 \\
 2  & 272 & 16 & 1.77 & 1.61 \\
 2  & 272 & 32 & 1.76 & 1.56 \\ 
 2  & 288 & 16 & 1.68 & 1.67 \\
 2  & 288 & 32 & 1.67 & 1.61 \\
\hline
\end{tabular}
\caption{Estimated and measured speed-up for a grid with $N=512$ points and 
  decomposition into 2 and 3 domains with two different overlaps.}
\label{speed}
\end{table}
Compared to the one dimensional case, 
a considerably higher speed-up is expected if decomposition is applied to
two dimensional problems, since it reduces the size of matrices 
to be inverted, the latter being a $N^2$ operation in the worst case for 
iterative methods and the matrices considered.

\section{Application to a P Cygni model}
\label{PCygni}
\begin{figure*}
  \epsfig{file=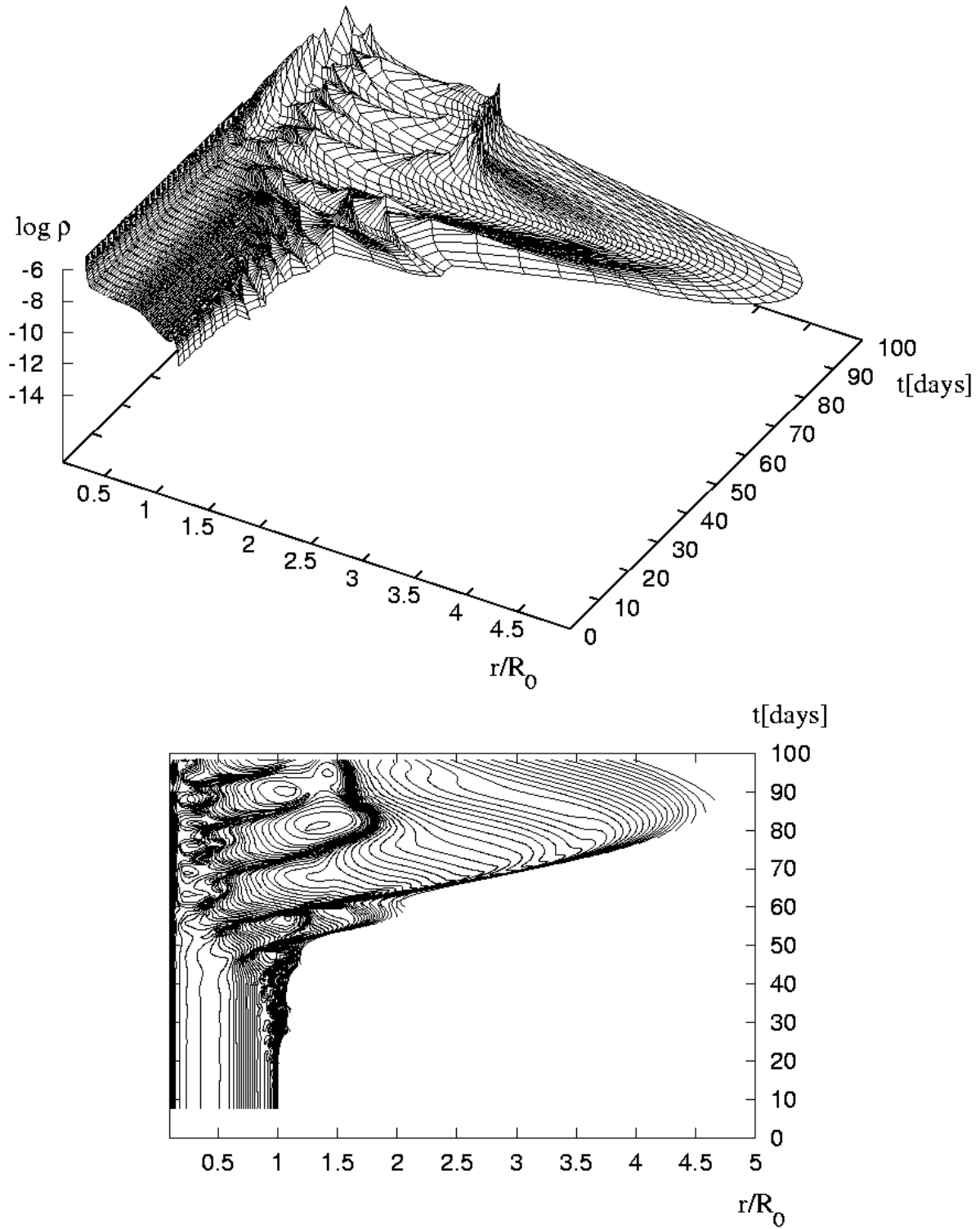,width=18cm}
  \caption{Density $\rho$ as a function of 
    radius (in units of the initial radius) 
    and time for an envelope  model of P Cygni   
    (top panel). The corresponding contour plot is given in the bottom
    panel. Strange mode instabilities, but no shock 
    oscillations are resolved in these diagrams.}
  \label{PCyg1} 
\end{figure*}
\begin{figure*}
  \epsfig{file=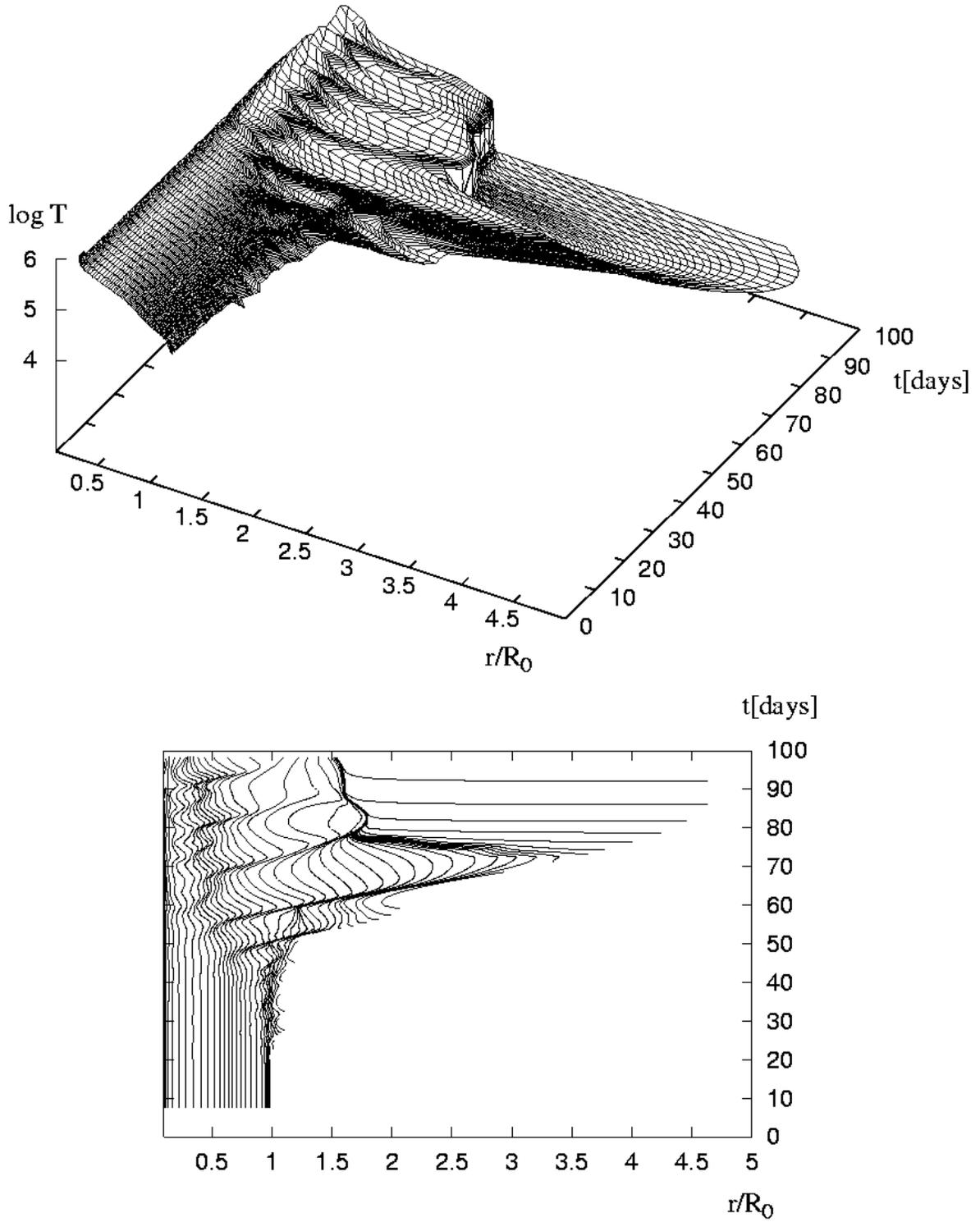,width=18cm}
  \caption{Same as Figure \ref{PCyg1}, but for the temperature $T$.}
  \label{PCyg2} 
\end{figure*}
\begin{figure*}
  \epsfig{file=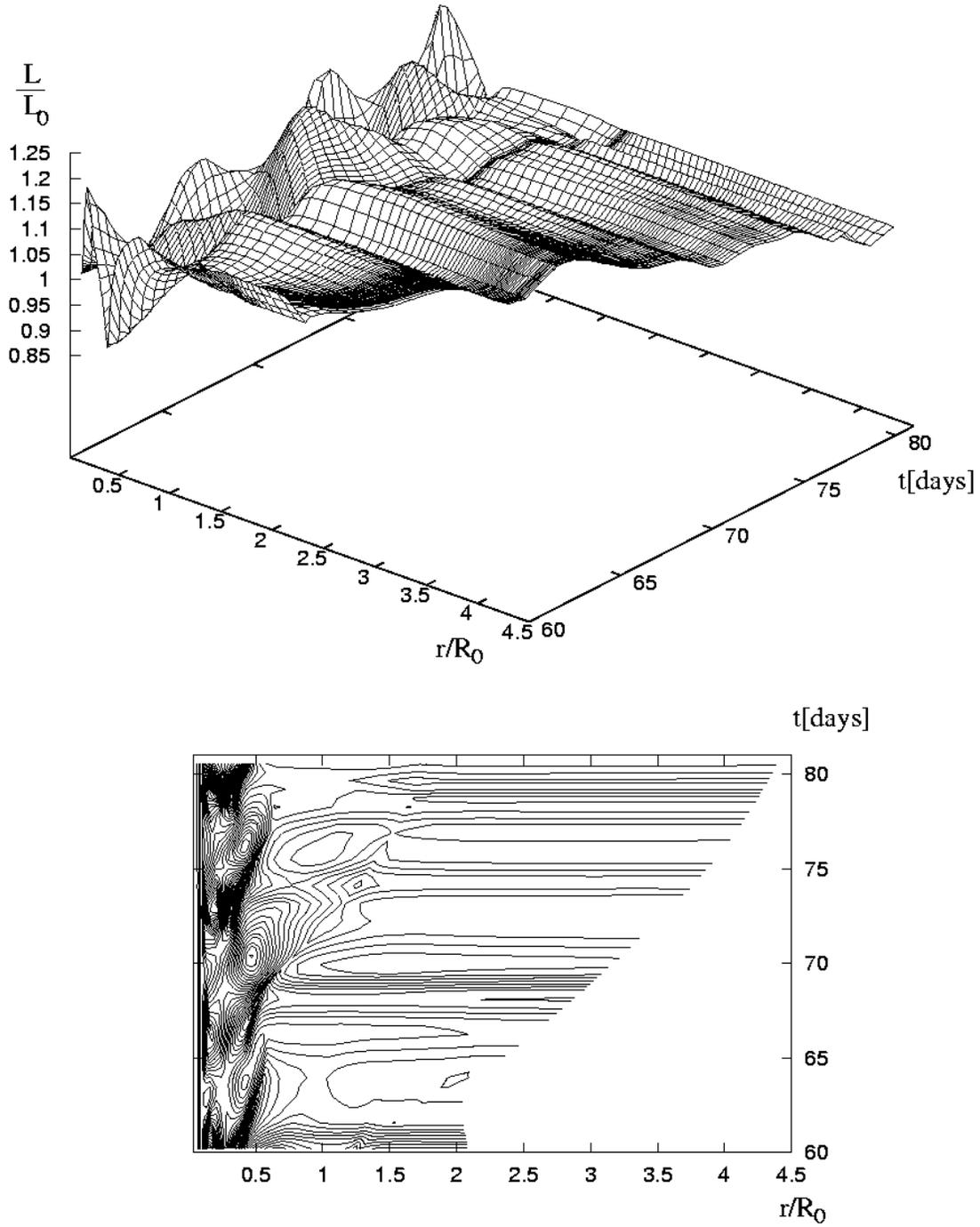,width=16.8cm}
  \caption{Luminosity in units of the initial
    luminosity as a function of 
    radius (in units of the initial radius) 
    before the shock capturing. The corresponding contour plot is
    given in the bottom panel. Note that the luminosity variations 
    take place on the dynamical timescale.}
  \label{PCyg5} 
\end{figure*}
\begin{figure*}
  \epsfig{file=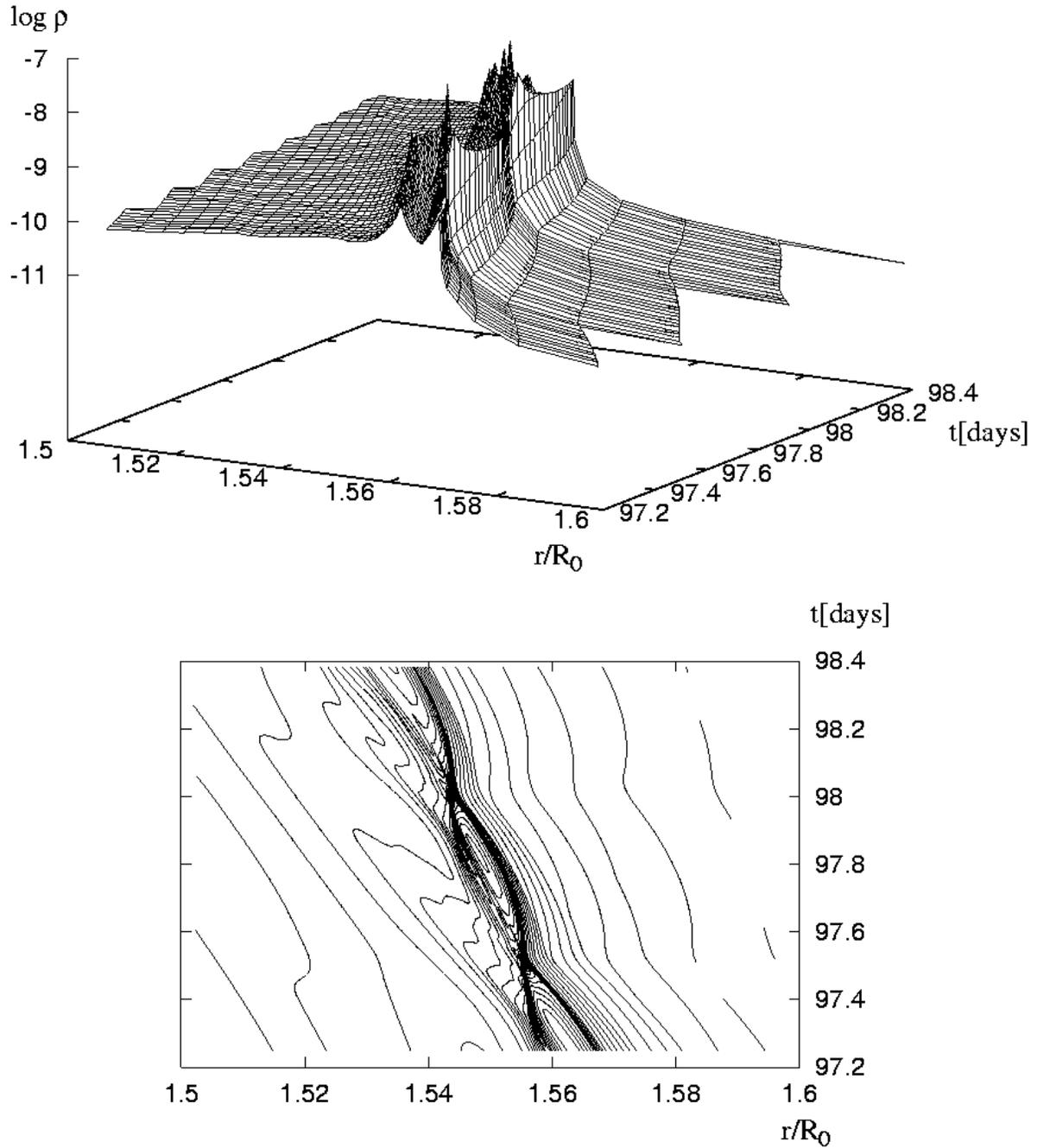,width=18cm}
  \caption{Same as Figure \ref{PCyg1} in the vicinity of the
    captured shock and for a time interval appropriate to  resolve
    the shock oscillations.}
  \label{PCyg3} 
\end{figure*} 
\begin{figure*}
  \epsfig{file=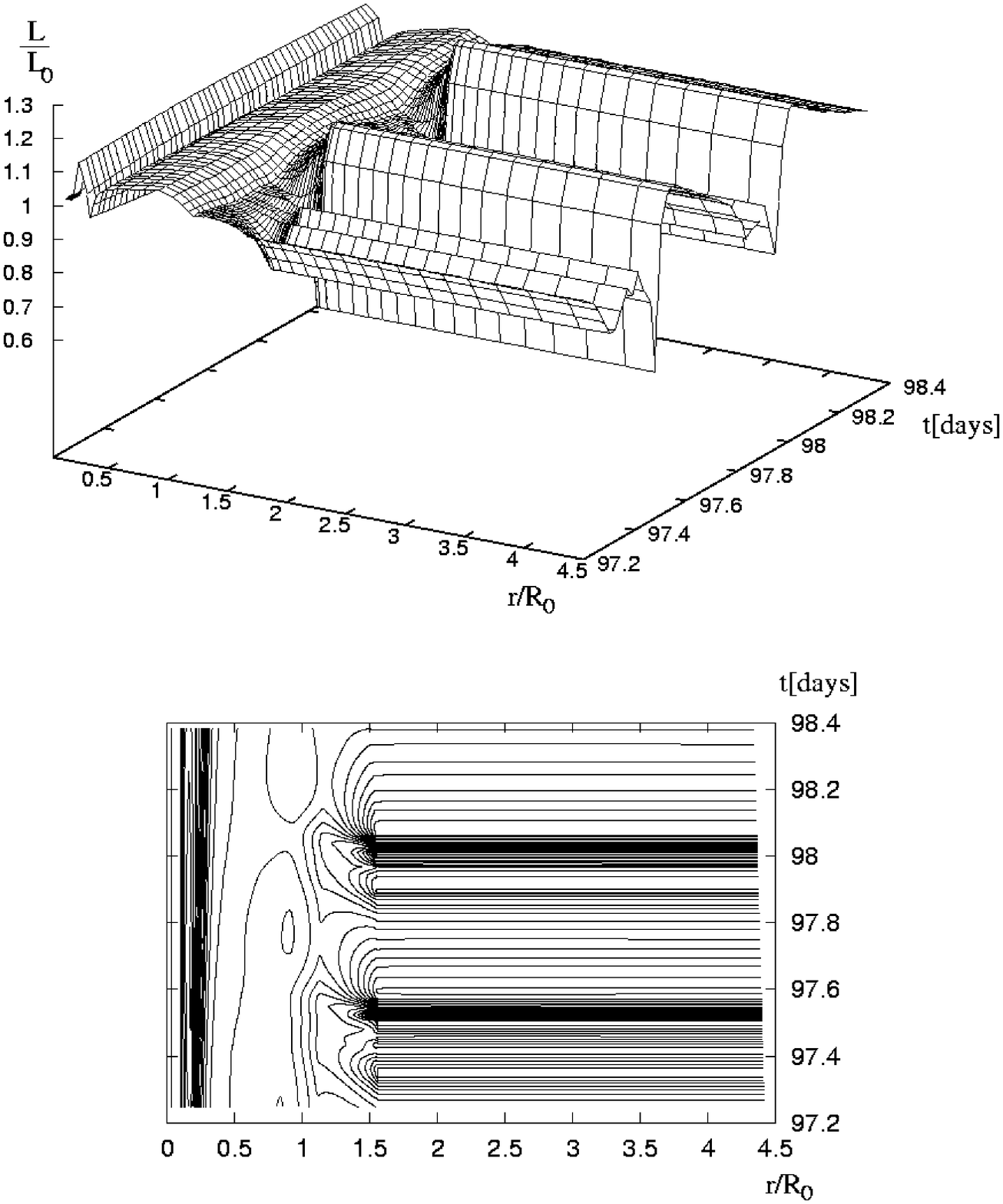,width=16.8cm}
  \caption{Luminosity in units of the initial
    luminosity as a function of 
    radius (in units of the initial radius) 
    and a time interval appropriate to  resolve
    the shock oscillations (top panel). The corresponding contour plot is
    given in the bottom panel. Note that the rapid photospheric 
    luminosity perturbations are
    defined at the captured shock (at $r/R_0\sim 1.5$). 
    Below the shock, luminosity perturbations
    are governed by strange mode instabilities operating on 
    the dynamical timescale.}
  \label{PCyg4} 
\end{figure*}

We apply the method discussed above to a stellar envelope model 
with parameters close to those observed for the star P Cygni.
Concerning luminosity, effective temperature and chemical composition
for this object, various authors (\cite{NH97} and \cite{PP90}) agree that
these parameters should lie in the vicinity of 
$L=752.5\cdot10^3L_{\odot}$, $T_{\rm{eff}}=19300K$ and  
$X=0.31$, $Y=0.67$, $Z=0.02$. The most uncertain parameter of P Cygni
is its mass. Standard stellar evolution calculations indicate a mass of
$M=50M_{\odot}$ (\cite{EH93}), whereas spectroscopic observations are consistent
with masses down to  $M=23M_{\odot}$ (\cite{PP90}). For our simulations 
we adopt $M=26.5M_{\odot}$, a value supported by the observation.  
Even with a more conservative mass of $M=50M_{\odot}$, the model 
is known to be unstable with respect to strange modes (\cite{KFG93}).
The higher value of $L/M$ adopted here amplifies the tendency towards
instability through shorter thermal timescales.

The simulation of strange mode instabilities in P Cygni starts
from the envelope model in hydrostatic equilibrium without any external
perturbations. Strange mode instabilities develop from numerical noise,
pass the linear regime of exponential growth and form multiple shocks
in the non-linear domain. One of these shocks is captured in the 
H-ionisation zone and starts oscillating on timescales of the order of the
sound travel time across the front ($\sim$ 0.5 days),
whereas its mean position varies on the dynamical timescale ($\sim$ 10 days).
In this phase of the evolution on two different timescales 
domain decomposition is used to speed up the calculation considerably.   

Figure \ref{PCyg1} shows the density $\rho$ as a function of 
radius (in units of the initial radius) and time for the envelope model of P Cygni.
The corresponding contour plot is given in the bottom panel. 
Note, that contour lines here and in the following are not always 
closed, since during the evolution of the star e.g. the density drops
to values lower than those of the initial model. 
After reaching the
non-linear regime at $t=\sim 20$ days, shocks are formed in the outer envelope,
travelling outwards and inflating the envelope successively up to 4.5 initial radii.
During this period, the surface velocity reaches 55 per cent of the escape
velocity at $t=75$ days.
After $\sim 80$ days the envelope starts to collaps, and a shock originating at
$r/R\approx 0.6$ at time $t=70$ days is then captured in 
the H-ionisation zone around $r/R\approx 1.5$ at $t\approx 85$ days. Subsequent
shocks, generated by the primary strange mode instability, 
are confined to the region below the captured shock.

Figure \ref{PCyg2} is the analogue to Figure \ref{PCyg1} for the temperature $T$.
The shocks discussed above can easily be identified in the contour plot. Note the
steep temperature gradient at the captured shock after its formation. 

Figure \ref{PCyg5} shows the luminosity in units of the initial
luminosity as a function of radius (in units of the initial radius) and 
a time interval before the shock capturing. The  luminosity varies on the 
dynamical timescale by 10 per cent, corresponding to a bolometric 
variation of $\sim 0.1^{\rm{m}}$.
It is defined in the inner envelope and remains constant with respect to 
radius above. There, luminosity perturbations cannot be sustained due 
to the low specific 
heat and the associated short thermal timescales. These luminosity variations
remind of the microvariations in P Cygni, described by, e.g., \cite{GS01}.
These authors report on cyclic behaviour of the visual brightness
with amplitudes of  
$\sim 0.1^{\rm{m}}$ and cycle lengths between 10 and 25 days, best fitting  
a quasi-period of $17.3^{\rm{d}}$. 

Figure \ref{PCyg3} shows the density $\rho$ as a function of 
radius (in units of the initial radius) in the vicinity of the 
captured shock and a time appropriate to resolve the shock oscillations. 
In particular, the corresponding contour plot in the bottom panel 
shows variation on two
different timescales. It exhibits both shock oscillations 
with a mean period of $\sim$0.5 days and the evolution on the
dynamical timescale, as indicated by the variation of the 
shock position between $r/R_0\approx 1.56$ and $r/R_0\approx 1.54$.

Figure \ref{PCyg4} shows the luminosity in units of the initial
luminosity as a function of radius (in units of the initial radius) and 
a time interval apropriate to resolve the shock oscillations. 
The rapid photospheric luminosity perturbations are
defined at the captured shock (at $r/R_0\sim 1.5$) and remain constant above it
(see discussion of Figure \ref{PCyg5}). The variation amounts to 20 per cent,
corresponding to $\sim 0.2^{\rm{m}}$  bolometrically.   
Below the shock, luminosity perturbations
are governed by strange mode instabilities operating on the dynamical timescale.
The photometric luminosity perturbation is therefore a superposition of two effects,
dominated by the fast oscillations induced by the shock.  
We emphasise, however, that the one dimensional calculations 
presented here have to be interpreted with
caution, since massive stars are known to suffer also from non-radial instabilities
(\cite{GM96}). Whether the captured shocks survive the deformations induced
by non-radial instabilities or become unstable themselves with respect to non-radial
perturbations, remains to be tested by at least two dimensional calculations.
They might break apart and could then contribute to the entire variation
by stochastically adding high-frequency perturbations to
the cyclic perturbations on the dynamical timescale induced by strange mode
instabilities. 
On the other hand, from the observations discussed by \cite{GS01}, 
no indications for the stability properties of the captured shock 
found here can be inferred, since the
time resolution of the data is not sufficiently high ($\sim$ 
one measurement per day).

Another effect of the primary strange mode instability consists of a mass transfer 
from the instability region into the outer parts of the stellar envelope. 
Owing to the Lagrangian approach
chosen here, this implies a reduced spatial resolution in the inner part of 
the stellar envelope.
Simultaneously, grid points are concentrated around the captured shock. 
For reliable calculations, however, a high resolution of the instability 
region is indispensable. Otherwise, the physical strange mode instability 
generating acoustic energy and shock waves is suppressed. 
Thus, due to insufficient resolution, the simulation
discussed had to be stopped after $\sim$100 days of simulated time. 
To overcome the difficulty  grid reconstruction is necessary. A 
corresponding algorithm, consistently inserting and eliminating 
gridpoints during the calculation is currently being developed and will be 
commented on in a forthcoming publication.  
 
\section{CONCLUSIONS}

Motivated by the discovery of high frequency shock oscillations in pulsating
stars confined in a narrow region, we have developed a procedure for an 
efficient treatment of such phenomena. The integration domain is decomposed 
into several subdomains according to the various, vastly 
different timescales present in the configuration.
To save computing time, these domains are integrated according to their
respective timescales. 
 
Criteria for the choice and decomposition of 
the computational domains have been derived. Decomposition implies artificial
inner boundaries which require suitable boundary conditions. How these have
to be chosen in order to minimise the numerical error (compared to the 
approach without decomposition) has been discussed. An overlap of the domains
was found to be necessary to produce satisfactory results.  

The decomposition technique has been tested and validated by comparison with
results obtained without decomposition. The major effect of decomposition
was found to consist of  a delay in time (phaseshift) with respect to the original
calculation. The latter is regarded as physically largely irrelevant.
Otherwise the numerical quality of the results was proven to be 
satisfactory. 
 
For the models considered,
decomposition was found both theoretically and by numerical tests to
reduce the computational costs by approximately 
a factor of two. The speed-up critically depends on the size of the
rapidly varying domain. Although a reduction of computing time by 50
per cent sounds moderate, it is of practical relevance considering the
duration of several weeks for a complete simulation.
The intended extension of the procedure to two dimensional problems
will yield an appreciably higher speed-up than that for the 
one dimensional model considered here, since the iterative inversion 
of some matrices there requires of the order of $N^2$ operations. Moreover,
decomposition in two dimensions may reduce the size of matrices to be inverted
such that fast direct methods (requiring of the order of $N$ operations) become
most efficient, whose application would then imply a further 
acceleration of the computation. 

We have applied the method to a model for the star P Cygni, 
paying special attention to
the adequate treatment of the different timescales involved. 

\section*{Acknowledgments}

We thank Professors K.J. Fricke and G. Warnecke for encouragement and support.
Financial support by the Graduiertenkolleg 
``Str\"omungsinstabilit\"aten und Turbulenz'' (MG) and 
by the DFG under grant WA 633 12-1 (SC) is gratefully acknowledged. The
numerical computations have been carried out using the facilities  
of the GWDG at G\"ottingen and of the ZIB at Berlin.

\bsp

\label{lastpage}

\end{document}